\documentclass[twocolumn,superscriptaddress,showpacs,preprintnumbers,amsmath,amssymb,prl]{revtex4}
\usepackage[T1]{fontenc}
\usepackage{graphicx}
\usepackage{dcolumn}
\usepackage{bm}

\begin{document}

\title{Strong coupling of light with one-dimensional quantum dot chain: from \\ Rabi oscillations to Rabi waves }

\author{G. Ya. Slepyan}%
\author{Y. D. Yerchak}
\email{Jarchak@gmail.com}
\author{S. A. Maksimenko}

\affiliation{%
Institute for Nuclear Problems, Belarus State University,
Bobruiskaya 11, 220050 Minsk, Belarus
}%
\author{A. Hoffmann}
\affiliation{Institut f\"{u}r Festk\"{o}rperphysik, Technische
Universit\"{a}t Berlin, Hardenbergstrasse 36, 10623 Berlin, Germany}


\begin{abstract}
Interaction of traveling wave of classic light with 1D-chain of
coupled quantum dots (QDs) in strong coupling regime has been
theoretically considered. The effect of space propagation of
Rabi oscillations in the form of traveling waves and wave packets
has been predicted. Physical interpretation of the effect has been
given, principles of its experimental observation are discussed.
\end{abstract}

\maketitle

\textit{Introduction.} -- Rabi oscillations are periodical
transitions of a two-state quantum system between its stationary
states in the presence of an oscillatory driving field, see e.g.
\cite{Scully}. First observed on nuclear spins in radio-frequency
magnetic field \cite{Torrey_1949}, the Rabi oscillations then were
discovered in many other two-level systems, such as atoms exposed to
electromagnetic wave \cite{Hocker_Tang_1968},  semiconductor QDs
\cite{Kamada_2001}, Josephson qubits \cite{Blais_07}, spin-qubits
\cite{Burkard_06}, and between ground and Rydberg atomic states
\cite{Johnson_08}. Besides the fundamental interest, the effect of
Rabi oscillations is promising for realization of binary logic and
optical control in quantum informatics and quantum computing.

Complication of physical systems where Rabi effect is observed
imposes  additional features on the ideal picture \cite{Scully} of
this effect. They are the time-domain modulation of the field-matter
coupling constant \cite{Law_96,yang_04}, the phonon-induced
dephasing \cite{forstner_03, Vagov_07} and the local-field effect
\cite{Slepyan_04,Paspalakis_06,Slepyan_07} -- just to mention a few.
New capabilities appear in systems of two coupled Rabi oscillators
\cite{Unold_05,Gea-Banacloche_06,Huges_05,Danckwerts_06,Ho_Trung_Dung_02,Salen_08,Tsukanov_06}.

In spatially extensive samples comprising a great number of
oscillators, the mechanism giving rise to Rabi oscillations induces
also a set of nonstationary coherent optical effects, such as
optical nutation, photon echo, self-induced transparency, etc.
\cite{Shen_84}. This is because the sample size exceeds
significantly wavelength and propagation effects come into play.  In
low-dimensional systems propagation effects are also manifested but
their character changes qualitatively. For example, the
computational model of the coherent intersubband Rabi oscillations
in a sample comprising 80 AlGaAs/GaAs quantum wells
\cite{Waldmueller_06} predicts the population dynamics to be
dependent on the quantum well position in the series. This result
demonstrates strong radiative coupling between wells and, more
generally, significant difference in the Rabi effect picture for
single and multiple oscillators. In the present Letter we build for the first time a
theoretical model of a distributed system of coupled Rabi
oscillators and predict the new physical effect: the propagation of Rabi
oscillations in space in the form of traveling waves and wave
packets.

\textit{Model and equation of motion.} -  Consider an interaction of
an one-particle excitation in an infinite periodical 1D chain of
identical coupled QDs with electromagnetic field. A $p$-th QD is
considered as a two-level system with $\left|b_p\right\rangle$ and
$\left|a_p\right\rangle$ as ground and excited states,
correspondingly, and the transition frequency $\omega_0$. Dephasing
and dissipation processes inside the QD are further neglected. The
coupling may originate from different physical processes (electron
tunneling, dipole-dipole interaction, etc.) and is accounted for in
the tight-binding approximation, i.e., is assumed to be restricted
to neighboring QDs. Let the QD chain be exposed to a plane wave
traveling along the chain, $E \sim \exp[-i(\omega t - kx)]$. The
wavenumber  satisfies the condition $ka \lesssim 1$ with $a$ as the
chain period, providing  later on the continuous limit transition.

In the one-particle basis, the Hamiltonian of the system  can be
represented by $\hat {H}=\hat {H}_0 +\Delta \hat {H}$, where the
term
\begin{equation}
\label{h_0}
\hat {H}_0 =
    \frac{\hbar \omega _0}{2} \sum\limits_n {\hat {\sigma }_{zn}}-
    \frac{\hbar\Omega _R}{2}\sum\limits_n {\hat {\sigma }_n^+
    e^{i(nka-\omega t)}+\mathrm{H.c.}}
\end{equation}
describes Rabi oscillations in non-interacting QDs. Here, $\hat
{\sigma }_{zn}$ and $\hat{\sigma }_n^+$ are the Pauli matrices for
$n$-th QD, and $\Omega _R $ is the real-valued Rabi frequency. The
term $\Delta \hat {H}$ accounts for the  QD-coupling and has the
form as follows:
\begin{equation}
\label{h_int} \Delta \hat {H}=-\frac{\hbar \xi}{2}\sum\limits_n
\sum\limits_{p=\pm 1}\left(\left|a_n \right\rangle \left\langle
a_{n+p} \right|+\left|b_n \right\rangle \left\langle b_{n+p} \right|
\right)+ \mathrm{H.c.},
\end{equation}
where $\xi$ is the coupling constant. Equation of motion has the
form of one-particle Schr\"{o}dinger equation $i \hbar
\partial_t\left| \Psi \right\rangle = \hat {H}\left| \Psi
\right\rangle$; the wave function can be written as $\left| {\Psi
(t)} \right\rangle =\sum_p \left(A_p(t) \left|a_p \right\rangle
+B_p(t) \left| b_p \right\rangle \right)$.  Taking into account
(\ref{h_0}) and (\ref{h_int}), we reduce the Schr\"{o}dinger
equation to a system of differential equations, which directly
couples  $A_p$ with $B_p, A_{p\pm1}$ and $B_p$ with $A_p,
B_{p\pm1}$. Carrying out then the continuous limit transition for
the variable $A_p(t)$ by $A_p(t)\rightarrow A(x,t)$, $A_{p+1}(t) +
A_{p-1}(t)-2A_{p}(t)\rightarrow a^2\partial^2_xA(x,t)$ and in the
same manner for the variable $B_p(t)$,  in the rotating-wave
approximation \cite{Scully} we arrive at the system of equations as
follows:
\begin{eqnarray}
\label{system1}
    \partial_t
    A=\displaystyle{-\frac{i}{2}(\omega_0-4\xi)A+\frac{i\Omega _R }{2}
    B e^{i(kx-\omega t)}+i\xi a^2\partial_x^2A}\,,
    \\ \rule{0in}{4ex}
    \partial_t B =
    \displaystyle{\frac{i}{2}(\omega_0+4\xi)B+\frac{i\Omega _R }{2}A
    e^{-i(kx-\omega t)}+i\xi a^2\partial_x^2B}\,,\label{system1a}
\end{eqnarray}
Eqs.  (\ref{system1}) and \eqref{system1a}  describe light -- QD
chain coupling  in the framework of formulated model.  Because we
are interested in the strong coupling regime, the quantity $\Omega
_R$ can not be considered as a small parameter and further analysis
of these equations is carried out without recourse to the
perturbation theory.

\textit{Traveling Rabi waves.} -- Let us consider elementary
solution of the system (\ref{system1})--\eqref{system1a} in the form
of traveling wave: $A \sim e^{i(h+k/2)x}e^{-i(\nu+\omega/2)t}$, $B
\sim e^{i(h-k/2)x}e^{-i(\nu-\omega/2)t}$, where $h$ is a given wave
number and $\nu$ is the eigenfrequency to be found. Solving
characteristic equation of the system
(\ref{system1})--\eqref{system1a} with respect to $\nu$  determines
the eigenfrequencies of system by
\begin{equation}
\label{disp_law} \nu_{_{1,2}} =
    \xi a^2h^2-\Phi\pm\left[\Bigl(\frac{\Delta}{2} +
    V h \Bigr)^2+\frac{\Omega_R^2}{4}\right]^{1/2},
\end{equation}
where $\Delta=\omega_0-\omega$ is the frequency detuning, $V = \xi k
a^2$ and $\Phi=2\xi(1-k^2a^2/8)$. These two solutions correspond to
two eigenmodes, given, respectively, by
\begin{equation}
\label{mode1}
 {\begin{array}{l}
A_1(x,t)=-\displaystyle\frac{ C_1\Omega _R
    e^{i(h+k/2)x}e^{-i(\nu_1+\omega/2)t}}
    {2(\nu_1+\Phi-\Delta/2-Vh-\xi a^2h^2)}\,, \\
    \rule{0in}{4ex}
B_1(x,t) = C_1 e^{i(h-k/2)x}e^{-i(\nu_1-\omega/2)t}\,,
\end{array}}
\end{equation}
and
\begin{equation}
\label{mode2} {\begin{array}{l} A_2(x,t) = C_2
    e^{i(h+k/2)x}e^{-i(\nu_2+\omega/2)t}\,,\\
    \rule{0in}{6ex}
B_2(x,t)=-\displaystyle\frac{C_2\Omega_R
     e^{i(h-k/2)x}e^{-i(\nu_2-\omega/2)t}}{2(\nu_2+\Phi+\Delta/2+Vh-\xi
a^2h^2)}\,,
\end{array}}
\end{equation}
where $C_{1,2}$ are normalizing constants.  Either of these modes is
a \textit{superposition} of ground and excited states, whose partial
amplitudes \textit{oscillate both in time and space}. Binding of
ground and excited states is caused by interaction of light with QD
chain and vanishes in the limit of $\Omega_R \rightarrow 0$. In that
case, Eqs. (\ref{mode1}) and (\ref{mode2}) describe QD-chain
excitons in equilibrium and inverse states, respectively. Retaining
in the expansion terms  linear in $|\Omega_R|$  and simultaneously
substituting $\Delta\rightarrow\Delta-i0$ we arrive at the
intermediate case of excitons weakly coupled  with electromagnetic
field. 

Space oscillations of the partial amplitudes are due to QD-coupling
and vanishes in the limit of $\xi \rightarrow 0$. Thus, each of
these modes can be interpreted as a Rabi wave with the frequency
determined by Eq. (\ref{disp_law}). In general case these waves are
excited simultaneously, while any of them can be excited separately by
a proper choice of initial conditions.

\begin{figure}[htb]
\includegraphics[width=0.48\textwidth,trim=15 15 15 20,clip]{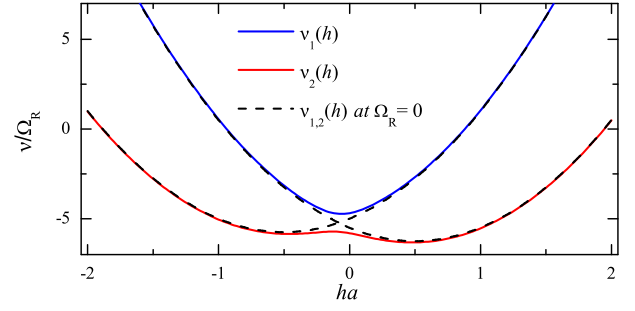}
\caption{\label{fig1} Typical dispersion  law for Rabi waves. $\xi =
3 \Omega_R$,  $\Delta = 0.5 \Omega_R$,  $ka=1$. Turndown of $ha$
corresponds to the range of applicability of the continuous limit
model.}
\end{figure}
Typical dispersion characteristics of Rabi waves are depicted in
Fig. \ref{fig1}. It should be noted that at given amplitude and
frequency of external field their frequencies have continuous
spectra (wave number $h$ varies continuously). Dispersion
dependences depicted in the figure are asymmetric:
$\nu_{1,2}(h)\neq\nu_{1,2}(-h)$. Physically, this is due to the
presence of preferential direction, which is determined by the
direction of light propagation along the chain. The QD coupling
leads to the inequality $\nu_1\neq-\nu_2$. That is why, unlike to
single QD, in QD chains the inversion oscillates anharmonically.
These oscillations can be represented as amplitude-modulated
harmonic oscillations with the frequency $(\nu_1-\nu_2)/2$, while
the modulation frequency is given by $(\nu_1+\nu_2)/2$. In traveling
Rabi wave, the inversion is constant in space.  Because $\nu_{1,2}(h)$ are real 
at any real $h$, the system is stable \cite{Landau_Lifshitz_1981}. Now, let us analyse the dispersion characteristics $h_{1,2}(\nu)$, assuming the frequency $\nu$ to be a given parameter. It is seen from Fig.\ref{fig1} that in some frequency range $h_1$ is complex and $h_2$ is real for real $\nu$. It corresponds to non-transmission of the first Rabi wave \cite{Landau_Lifshitz_1981}. The frequency range in which both of $h_{1,2}$ are complex also exists. This case corresponds to complete non-transmission of the Rabi waves with given frequency.

Note that the eigenmodes (\ref{mode1}) and  (\ref{mode2}) each
comprise traveling waves with \textit{different} wave numbers $h \pm
k/2$. Physically, this means that the Rabi wave propagates in an
effective periodically inhomogeneous medium formed by spatially
oscillating (with period $2\pi/k$) electric field. Therefore, the
diffraction is developed in the system.  In the limit $k \rightarrow
0$ the medium turns homogeneous and the diffraction effect vanishes.

Assuming the frequency $\nu$ to be a given parameter and solving Eq.
(\ref{disp_law}) with respect to wave number, we  obtain for $k = 0$:
\begin{equation}
\label{disp_law2} h_{1,2} =
\pm\left[\frac{1}{a^2\xi}\Bigl(\nu+2\xi\pm\frac{1}{2}\sqrt{\Omega
_R^2+\Delta^2}\,\Bigr)\right]^{1/2},
\end{equation}
where external signs  $\pm$ correspond to two directions of
propagation while signs  $\pm$ before internal radical correspond to
two types of Rabi waves indexed by 1 and 2. It directly follows from
(\ref{disp_law2}) that electric field forms an effective medium for
propagating Rabi waves. In inhomogeneous electric field the Rabi
frequency becomes coordinate-dependent, $\Omega _R = \Omega _R(x)$,
and therefore the medium becomes inhomogeneous too providing
reflection of Rabi waves and their mutual transformations at the
inhomogeneities. In that way one obtain a unique ability to control
the processes of the reflection and dispersion of Rabi waves by
varying the light spatial distribution. As a potential realization
scheme we indicate the interaction of QD-chain with Gaussian light
beam (or superposition of such beams) with the beams' widths and
mutual disposition as controllable factors.

\begin{figure}[ht]
\includegraphics[width=0.48\textwidth,trim=35 15 28 25,clip]{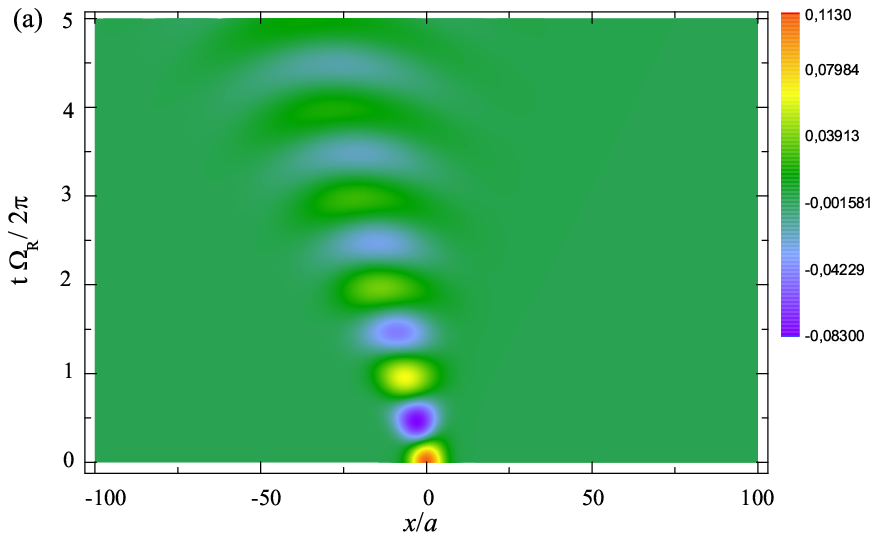}
\includegraphics[width=0.48\textwidth,trim=35 15 28 25,clip]{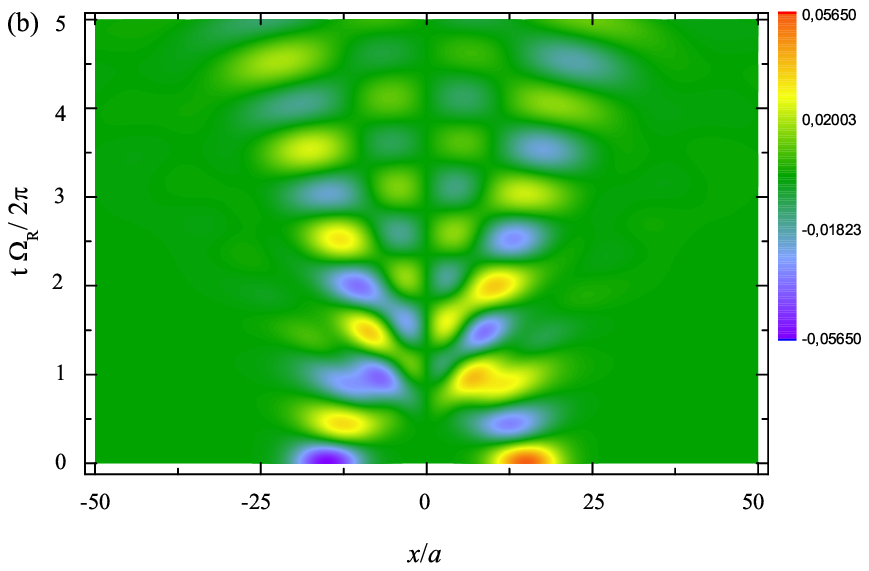}
\caption{\label{fig2} Space-time distribution of the inversion in
the QD chain (color online). a) A single Gaussian wavepacket $A(x,0)
= \exp(-x^2/2\sigma^2)/\sqrt[4]{\pi\sigma^2}$, $B(x,0)=0$, $\Delta =
Vk$. b) Two counterpropagating identical Gaussian wavepackets:
$A(x,0) = \exp[-(x-3\sigma)^2/2\sigma^2]/\sqrt[4]{4\pi\sigma^2}$,
$B(x,0) = \exp[-(x+3\sigma)^2/2\sigma^2]/\sqrt[4]{4\pi\sigma^2}$,
$\Delta = 0$.  In both cases
$ka=0.33$, $\sigma = 5 a$,  $\xi = 3 \Omega_R$.}
\end{figure}

\textit{Rabi wave packets.} - The process of the excitation
transition opens up new opportunities for controlling the dynamics
of Rabi oscillations. For identification of control factors we need
to know general solution of the system
(\ref{system1})--\eqref{system1a}. To find it, we first introduce
the variables $u(x,t) = A(x,t) \exp\left[{i}(\omega t - kx - 2\Phi
t)/{2}\right]$, $v(x,t) = B(x,t) \exp\left[-{i}(\omega t - kx +
2\Phi t)/{2}\right]$. For these variables, Eqs.
(\ref{system1})--\eqref{system1a} are reduced to the form as
follows:
\begin{eqnarray}
\label{system2} \displaystyle{\partial_t u +\frac{i \Delta}{2}u+V
\partial_xu-i\xi a^2\partial^2_xu-\frac{i\Omega _R }{2}v =0\,,}
 \rule[-5mm]{0mm}{1mm}\cr \displaystyle{\partial_t v -\frac{i \Delta}{2}v-V
\partial_xv-i\xi a^2\partial^2_xv-\frac{i\Omega_R }{2}u
=0\,.}\nonumber
\end{eqnarray}
This system can be solved exactly by using the Fourier transform
with respect to $x$. Finally we arrive at
\begin{eqnarray}\label{sol}
u(x,t)=\int\limits_{-\infty }^\infty\!\!
    {\left[{\tilde{u}(h)\varphi^-_h(t)+\tilde{v}(h)\psi_h(t)} \right]
    e^{ih(x-\xi a^2ht)}} dh,
\\
v(x,t)=\int\limits_{-\infty }^\infty\!\!
    {\left[{\tilde{u}(h)\psi_h(t)+\tilde{v}(h)\varphi^+_h(t)}
\right]e^{ih(x-\xi a^2ht)}} dh,\,\,\label{sol1}
\end{eqnarray}
where $\varphi ^\pm _h (t)=\cos \tau \pm i(\Delta_h /\Omega_h )\sin
\tau$, $\tau ={\Omega_h t}/{2}$, $\psi_h (t)=i({\Omega _R
}/{\Omega_h })\sin \tau$, $\Omega_h =\sqrt {\Omega_R^2+\Delta_h^2}$,
$\Delta_h = \Delta + 2Vh$\,. The function $\tilde{u}(h)$ is
determined by the initial condition
\begin{equation}
\tilde {u} (h)=\frac{1}{2\pi }\int\limits_{-\infty }^\infty {u(x,0)e^{-ihx}} dx
\end{equation}
and the function  $\tilde{v}(h)$ -- analogously. It is easy to
verify, that the solution obtained satisfies the probability
conservation law
$\int_{-\infty}^{\infty}[|u(x,t)|^2+|v(x,t)|^2]\,d\,x=1$ for any
$t>0$. A typical space-time distribution of the spatial density of
the  inversion  (the inversion per single QD)
$w(x,t)=a[|u(x,t)|^2-|v(x,t)|^2]$
 is shown in Fig. \ref{fig2}a.
As follows from Eqs. \eqref{sol}-\eqref{sol1} the depicted space-time dynamics of
$w(x,t)$ corresponds to the evolution of a Rabi wavepacket defined
as a superposition of Rabi waves with continuous spectrum of
$\Omega_h$. Physical interpretation of the picture predicted to
observe in the QD chain can be given on base of the
collapse-revivals concept \cite{Scully}. Distinctive feature of the
case considered is that the distribution of collapses and revivals
is permanently varied in space and time.

\begin{figure}[ht]
\includegraphics[width=0.48\textwidth,trim=15 17 15 10,clip]{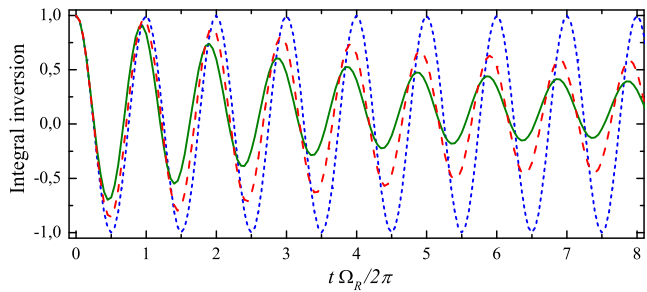}
\caption{\label{fig3} Temporal dependence of the  integral inversion
at the iтput parameters as follows:  $\Delta = 0$, $k = 0$ (dotted
line); $\Delta = 0$, $ka = 0.33$ (solid line); $\Delta = Vk$, $ka =
0.33$ (dashed line). In all cases, $\xi = 3 \Omega_R$ and $\sigma =
5a$. }
\end{figure}

Although  variation of the inversion density, depicted in Fig.
\ref{fig2}, in arbitrary point of the space occupied by the Rabi
wavepacket is not too large, an integral characteristics presented
in Fig. \ref{fig3} -- the ''integral'' inversion
$\int_{-\infty}^{\infty}w(x,t)dx$ -- of initially unpolarized
QD-chain oscillates between -1 and 1, thus indicating presence of
strong light-QD coupling.

Note that oscillations of the integral inversion at $V \neq 0$ damp
with time (see Fig. \ref{fig3}), whereas such a damping is absent at
$V = 0$ and integral inversion oscillates harmonically in the range
from -1 to 1 (dotted curve in Fig. \ref{fig3}). Such a behaviour
indicates appearance of a specific mechanism of collective
dephasing. Physically, this is because the effective detuning
$\Delta_h$ and therefore the carrier frequency  $\Omega_h$ of Rabi
oscillations constituting the wavepacket depends on $h$ (Doppler
shift). As different from that, the condition $V = 0$ keeps the
$h$-dependence only in the amplitude modulation frequency
$(\nu_1+\nu_2)/2$ and thus does not result in dephasing. In the weak
coupling limit the indicated dephasing mechanism is analogous to the
Landau damping in plasma.

The dependence of the frequency of Rabi oscillations $\Omega_h$  on
$V$ shows that the value $\Delta =0$ is not optimal for the effect
observation. Optimization of $\Delta$ allows increasing the
intensity of Rabi wave and the dephasing time, see dashed curve in
Fig. \ref{fig3}. In the frequency domain, fine tuning of the system
to the resonance is achieved by the variation of $V$ (changing the
angle of incidence of light).

Interaction  of two counterpropagating identical Gaussian Rabi
wavepackets elastically colliding at  $x=0$ is shown in Fig.
\ref{fig2}b.  Although the inversion oscillates in time and moves in
space, integral inversion of initially saturated QD chain
($\int_{-\infty}^{\infty}w(x,0)dx=0$) does not experience
oscillations: this quantity equals zero for all $t\geq 0$ and
arbitrary values of $\Omega_R$.

\textit{On experimental observability of Rabi waves.} -  The theory
presented can be extended to a various physical situations, such as
quantum dynamics of an electron in QD chain. Rabi waves are realized
via optically induced transitions between the size-quantized
electron levels. Transition of electron from one QD into another
occurs by means of tunneling through the potential barrier
\cite{ref02, Tsukanov_06}. Another example is the Rabi oscillations
of two-electron entangled state $(|01\rangle+|10\rangle)/\sqrt{2}$
taking place in two neighboring QDs due to dipole-dipole interaction
\cite{Unold_05, Gea-Banacloche_06, Huges_05, Danckwerts_06,
Ho_Trung_Dung_02, Salen_08}. The model developed describes the
motion of this two-electron state as a single whole; in this case
the wavefunction $\left|\Psi\right\rangle$ is the envelope function.

Theoretical analysis carried out has shown that the optimal
excitation of Rabi waves require the coupling factors of both
neighboring QDs and single QD with field to be comparable by
magnitude: $\xi \sim \Omega_R$. Another necessary condition being
imposed on the Rabi frequency is essential exceeding  over all
intrinsic relaxation rates. For typical QD structures
\cite{Kamada_2001}, this condition is satisfied for Rabi frequency
varied over a wide range $\hbar\Omega_R \sim 0.001-1$ meV. This
range corresponds to realistic electric field variation $E \sim 10^2
- 10^5$ V/cm. As a consequence, the interdot coupling constant $\xi$
also varies between  1 $\mu$eV and 1 meV what is practically
achievable \cite{Unold_05,Tsukanov_06, Danckwerts_06} for typical
interdot distances $4 - 20$ nm.

Experimentally, the Rabi waves can be detected in resonant
fluorescence spectra of spatially extensive samples by the presence
of  \textit{new spectral lines}, additional to the Mollow triplet, as well as by the Doppler shift and broadening of the triplet lines, etc. Of course, highly ordered chains of
uniform QDs are required to exclude nonhomogeneous broadening, which
may hide the effect.  Impressive progress in growing of perfect
nanostructured successions achieved in last years (e.g., see
\cite{Mano}) is very promising for that aim.

Rabi waves can be observed in systems of another physical nature
such as quantum electrical circuits \cite{Blais_07}, if one proceed
from two coupled Josephson cubits in microstrip resonator
\cite{Blais_07} to a distributed structure of such elements imposed
to interqubit interaction. In particular, in that structure the Rabi
wave frequency goes down to microwaves and the field intensity
necessary for Rabi waves excitation decreases.

\textit{Conclusion.} -  In this Letter, we have predicted the
exitance of Rabi waves -- wave propagation of the inversion in
spatially extensive systems of coupled oscillators. The system has
been exemplified by an 1D-chain of coupled QDs exposed to an
intensive traveling light wave.  Spatial propagation of Rabi
oscillations in the form of traveling waves and wave packets is
shown to occur in the chain. The propagation is predicted to be
accompanied by the damping of Rabi wave in time manifesting new
mechanism of collective dephasing, which is an analog of Landau
damping of exciton-polaritons extended to the strong light-matter
coupling.

Authors acknowledge a support from the INTAS (grant
05-1000008-7801). The work of S.A.M. was partially carried out
during the stay at the Institut f\"{u}r Festk\"{o}rperphysik, TU
Berlin, and supported by the Deutsche Forschungsgemeinschaft (DFG).
Authors are grateful to Dr. J. Haverkort for stimulative
discussions.


\begin{thebibliography}{99}
\bibitem{ref02}  The tunnel transparency of the barrier is not always identical for both states. As this occurs,
the coefficients at $\left|a_n\right\rangle\left\langle
a_{n+p}\right|$ and $\left|b_n\right\rangle\left\langle
b_{n+p}\right|$  in (\ref{h_int}) turn out to be different. The
theory developed can easily be extended to this case. Corresponding
calculations does not show substantial change of the Rabi waves
dynamics comparing with that presented in Fig.\ref{fig2}a.
\end{thebibliography}
\end{document}